\documentclass[10pt,journal]{IEEEtran}

\usepackage{nomencl}
\makenomenclature

\renewcommand{\IEEEQED}{\IEEEQEDclosed}
\usepackage{color}

\usepackage{subfigure}
\usepackage{graphicx,cite,epsfig,amssymb,amsmath,multirow,lettrine,flushend,extarrows}
\usepackage{algorithm}
\usepackage{algorithmic}

\begin{document}
%
\title{Distributed Real-Time HVAC Control for Cost-Efficient Commercial Buildings under \\ Smart Grid Environment}
\author{{Liang~Yu,~\IEEEmembership{Member,~IEEE}, Di Xie, Tao~Jiang,~\IEEEmembership{Senior Member,~IEEE}, \\Yulong Zou,~\IEEEmembership{Senior Member,~IEEE},~and Kun~Wang,~\IEEEmembership{Senior Member,~IEEE}}
\thanks{
\newline L. Yu, D. Xie, Y. Zou, and K. Wang are with Key Laboratory of Broadband Wireless Communication and Sensor Network Technology of Ministry of Education, Nanjing University of Posts and Telecommunications, Nanjing 210003, P. R. China.  \newline
T. Jiang is with Wuhan National Laboratory for Optoelectronics, School of Electronics Information and Communications, Huazhong University of Science and Technology, Wuhan 430074, P. R. China.\newline

}}

\markboth{IEEE Internet of Things Journal,~Vol.~XX, No.~XX, Month~2017}%
{Liang \MakeLowercase{\textit{et al.}}: Distributed Real-Time HVAC Control for Cost-Efficient Commercial Buildings Under Smart Grid Environment}

\maketitle

\begin{abstract}
In this paper, we investigate the problem of minimizing the long-term total cost (i.e., the sum of energy cost and thermal discomfort cost) associated with a Heating, Ventilation, and Air Conditioning (HVAC) system of a multizone commercial building under smart grid environment. To be specific, we first formulate a stochastic program to minimize the time average expected total cost with the consideration of uncertainties in electricity price, outdoor temperature, the most comfortable temperature level, and external thermal disturbance. Due to the existence of temporally and spatially coupled constraints as well as unknown information about the future system parameters, it is very challenging to solve the formulated problem. To this end, we propose a realtime HVAC control algorithm based on the framework of Lyapunov optimization techniques without the need to predict any system parameters and know their stochastic information. The key idea of the proposed algorithm is to construct and stabilize virtual queues associated with indoor temperatures of all zones. Moreover, we provide a distributed implementation of the proposed realtime algorithm with the aim of protecting user privacy and enhancing algorithmic scalability. Extensive simulation results based on real-world traces show that the proposed algorithm could reduce energy cost effectively with small sacrifice in thermal comfort.
\end{abstract}

\IEEEpeerreviewmaketitle

\section{Introduction}\label{s1}

The smart grid has been considered as one of the most important applications of Internet of Things (IoT) technologies in recent years, which aims to provide reliable, secure, and efficient energy delivery to consumers\cite{LiJIOT2017,XuJIOT2017}. As large consumers in smart grids, buildings consume a significant portion of electricity in a country\cite{PanJIOT2015}. For example, residential buildings and commercial buildings accounted for 38.7\% and 35.5\% of the total electricity usage of U.S. in 2010\cite{Book2011}, respectively. In commercial buildings (e.g., offices, stores, restaurants, warehouses, other buildings used for commercial purposes), HVAC (Heating, Ventilation, and Air Conditioning) systems account for about 45\% of the total electricity usage, which leads to high energy cost for the operators of commercial buildings. To reduce the energy cost of a commercial building, the direct way is to reduce the power input of HVAC systems, which would affect thermal comforts of occupants. Thus, it is very important to jointly manage the energy cost and thermal discomfort associated with HVAC systems in commercial buildings.

In this paper, we consider a commercial building with an HVAC system and multiple temperature zones. The purpose of this paper is to minimize the long-term total cost (i.e., the sum of energy cost and thermal discomfort cost) associated with the HVAC system under smart grid environment, where dynamic electricity prices could be exploited to save energy costs for electricity consumers\cite{Lei2012,LiangYuPowerOutage2015,Tao2014,LiangAccess2016,LiangIoT2016,Chai2016}. To achieve the above aim, we first formulate a stochastic program to minimize the time average expected total cost with the consideration of uncertainties in electricity price, outdoor temperature, the most comfortable temperature level, and external thermal disturbance. Due to the existence of temporally and spatially coupled constraints as well as unknown information about the future system parameters, it is very challenging to solve the formulated problem.

Typically, the framework of Lyapunov optimization techniques (LOT)\cite{Neely2010} is adopted to solve a time average optimization problem and a realtime energy management algorithm can be designed\cite{Guo2013,Fan2016}. A Lyapunov-based energy management algorithm intends to buffer the power demand requests of flexible loads in queues when electricity prices are high and to serve the stored requests when electricity prices are low. Different from flexible loads (e.g., electric vehicles) with specified energy/power demands, an HVAC system has unknown power demand that is related to many factors, namely the most comfortable temperature level decided by occupants, the lower and upper bounds of indoor temperature, outdoor temperature, and external thermal disturbance. Therefore, existing Lyapunov-based energy management algorithms can not be applied to our problem directly.

To avoid knowing about an HVAC power demand when using the LOT framework, we construct some virtual queues associated with indoor temperatures of all temperature zones. By stabilizing such queues and minimizing the total cost simultaneously, we can design a realtime algorithm without the need of predicting any system parameters. In addition, to protect user privacy and improve algorithmic scalability, we provide a distributed implementation for the proposed realtime algorithm. Theoretical analysis shows the feasibility and performance guarantee of the proposed distributed realtime algorithm. Moreover, simulation results based on real-world traces show the effectiveness of the proposed algorithm in the aspect of total cost reduction.

The main contributions of this paper are summarized as follows,
\begin{itemize}
  \item We formulate a stochastic program to minimize the time average expected total cost (i.e., the sum of energy cost and thermal discomfort cost) with the consideration of uncertainties in electricity price, outdoor temperature, the most comfortable temperature level, and external thermal disturbance.
  \item We propose a Cost-aware Distributed Realtime Algorithm (CDRA) to solve the formulated problem based on the LOT framework and binary search. CDRA does not require predicting any system parameters and knowing an HVAC power demand. Moreover, CDRA can protect user privacy and has good algorithmic scalability. In addition, we analyze the feasibility and performance guarantee of CDRA theoretically.
  \item Extensive simulation results based on real-world traces illustrate the effectiveness of CDRA, which can reduce energy cost effectively with small sacrifice in thermal comfort.
\end{itemize}

The rest of this paper is organized as follows. In Section \ref{s2}, we give the literature review. In Section \ref{s3}, system model and problem formulation are provided. Then, we propose a cost-aware distributed realtime algorithm in Section \ref{s4}. After conducting extensive simulations in Section \ref{s5}, we draw the conclusion and point out the future work in Section \ref{s6}.

\section{Literature Review}\label{s2}

Due to the high energy consumption of HVAC systems, the HVAC control in commercial buildings has attracted a lot of attention. Accordingly, various models and control methods have been developed to reduce energy consumption, energy cost, or thermal discomfort. For example, an HVAC control method based on MPC (Model Predictive Control) techniques was proposed in \cite{Mantovani2015} to reduce vertical thermal stratification and discomfort due to overheating in a commercial building. In \cite{Ma2015}, Ma \emph{et al.} presented a stochastic MPC-based HVAC control method to minimize the expected energy cost while bounding the probability of thermal comfort violations by exploiting stochastic information of weather and load learned from historical data. In \cite{Lin2015}, Lin \emph{et al.} conducted the experimental evaluation of frequency regulation from commercial building HVAC systems. In \cite{Vrettos2016}, Vrettos \emph{et al.} proposed a control framework for reliable provision of frequency reserves by aggregating HVAC systems of commercial buildings. In \cite{Hao2017}, Hao \emph{et al.} proposed a transactive control approach of HVAC systems in commercial buildings for demand response. In \cite{Radhakrishnan2017}, Radhakrishnan \emph{et al.} proposed a learning-based hierarchical distributed HVAC control method to minimize the energy consumption of a multizone commercial building with the consideration of some operational constraints, e.g., ventilation requirements. In \cite{Zhang2017}, Zhang \emph{et al.} designed a realtime distributed HVAC control strategy for a commercial building to minimize the weighted sum of energy consumption and thermal discomfort by solving a steady-state resource allocation problem.

Different from above-mentioned studies, we investigate the problem of minimizing the time average expected total cost (i.e., the sum of energy cost and thermal discomfort cost) associated with the HVAC system in a commercial building and propose a distributed realtime HVAC control method based on the LOT framework. The features of the proposed algorithm are summarized as follows: (1) without the need of predicting any system parameters and knowing their stochastic information; (2) protecting user privacy; (3) high algorithmic scalability. Though the LOT framework has been widely used in energy management of data centers\cite{LiangArxiv2016,Yu2015,LiangIoT}, microgrids\cite{Huang2014}, residential households\cite{Guo2013}, and smart homes\cite{Fan2016}, it is still not used in the HVAC control of commercial buildings. In \cite{Fan2016}, Fan \emph{et al.} investigated the online energy management problem for a smart home with an HVAC load based on the LOT framework. Specifically, this paper intends to minimize energy cost by buffering the power demand requests of appliances in queues when electricity prices are high and serving requests when electricity prices are low. However, different from loads with specific energy/power demands (e.g., electric vehicles), an HVAC load has unknown power demand that is related to many factors, such as the most comfortable temperature level decided by occupants, the lower and upper bounds of indoor temperature, outdoor temperature, and external thermal disturbance. Thus, the HVAC power demand is randomly generated in \cite{Fan2016} and can not reflect the true demand of the HVAC system. Though the LOT framework is also adopted to design the control method for an HVAC system in a commercial building, this paper has several aspects different from \cite{Fan2016}: (1) by constructing and stabilizing virtual queues associated with indoor temperatures of all temperature zones, our proposed algorithm operates without knowing the HVAC power demand; (2) we jointly consider the minimization of energy cost and thermal discomfort cost; (3) we consider the HVAC control in a multizone commercial building and the control decisions are the air supply rates of all zones, which are coupled with each other.

\section{System Model and Problem Formulation}\label{s3}

\begin{figure}[!htb]
\centering
\includegraphics[scale=0.59]{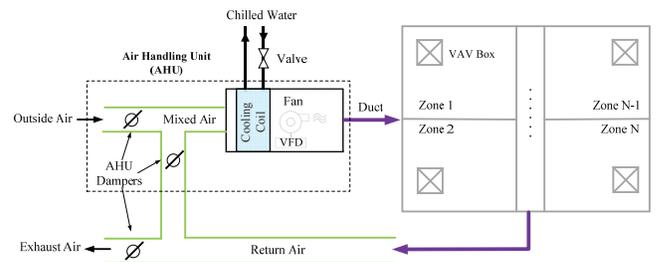}
\caption{Commercial HVAC system.} \label{fig_1}
\end{figure}

We consider a commercial building with $N$ zones (e.g., rooms), whose temperatures are adjusted by an HVAC system as shown in Fig.~\ref{fig_1}. To be specific, the HVAC system consists of an Air Handling Unit (AHU) for the whole building and a set of Variable Air Volume (VAV) boxes for each zone. The AHU is composed of dampers, a cooling coil, and a Variable Frequency Drive (VFD) fan. The dampers could mix the outside fresh air with the air returned from each zone to satisfy the ventilation requirement of each zone. The cooling coil cools down the mixed air and the VFD fan could deliver the mixed air to the VAV box in each zone. In each VAV box, there are a damper and a reheating coil, where the damper is used to adjust the rate of supply air and the reheating coil could reheats the supply air when needed. For simplicity, this paper mainly focuses on the case that all zones need cooling. Moreover, we ignore the heat transfer between neighboring zones similar to\cite{Hao2017,Radhakrishnan2017,Zhang2017}, since the total heat gain from the outside and the inside of a zone is (sometimes much) greater than that from neighboring zones. Note that the nature of the proposed algorithm would not change if the heat transfer between neighboring zones is considered. For easy understanding, we provide the main notations in Table~\ref{table_1}.

\begin{table}[!t]
\renewcommand{\arraystretch}{1.3}
\scriptsize
\caption{Notations}
\label{table_1}
\centering
\begin{tabular}{|c|l|}
\hline
 Symbol &  Definition \\
\hline\hline
$N$ & number of zones \\
\hline
$C_i$ & thermal capacitance of zone $i$  \\
\hline
$R_i$ & thermal resistance of the zone $i$  \\
\hline
$T_{i,l}$ & indoor temperature of zone $i$ at time $l$  \\
\hline
$T_{o,l}$ & outdoor temperature at time $l$  \\
\hline
$C_a$ & the specific heat of the air  \\
\hline
$m_{i,l}$ & air supply rate of zone $i$ at time $l$ \\
\hline
$T_s$ & air temperature of the supply fan  \\
\hline
$q_{i,l}$ & external thermal disturbance in zone $i$ at time $l$ \\
\hline
$\tau$ & time interval \\
\hline
$t$ & index of time intervals \\
\hline
$M$ & total number of time intervals \\
\hline
$T_i^{\min}$, $T_i^{\max}$ & minimum, maximum indoor temperature of zone $i$ \\
\hline
$m_i^{\min}$, $m_i^{\max}$ & minimum and maximum air supply rate of zone $i$ \\
\hline
$\overline{m}$ & upper bound of the total air supply rate  \\
\hline
$\phi_i$ & the cost coefficient related to thermal discomfort \\
\hline
$\mu$ & the coefficient related to fan power consumption\\
\hline
$S_t$ & electricity price at slot $t$  \\
\hline
$\gamma$ & the damper position in the AHU \\
\hline
$T^{\text{ref}}_{i,t+1}$ & the most comfortable temperature of zone $i$ at slot $t+1$ \\
\hline
$\eta$ & the efficiency factor of the cooling coil \\
\hline
$COP$ & coefficient of performance of the chiller\\
\hline
$Q_{i,t}$ & virtual queue associated with indoor temperature at zone $i$ \\
\hline
$\delta_i$ & the shifted parameter related to $T_{i,t}$ \\
\hline
$L_{t}$ & Lyapunov function at slot $t$  \\
\hline
$\Delta_t$ &  one-slot conditional Lyapunov drift \\
\hline
$\Delta Y_t$ &  \emph{drift-plus-penalty} term \\
\hline
\end{tabular}
\end{table}

\subsection{Commercial HVAC model}
For a temperature zone, its thermal dynamics could be described by the following model\cite{Hao2017},
\begin{align} \label{f_1}
C_i\frac{{d{T_{i,l}}}}{{dl}} = \frac{{{T_{o,l}} - {T_{i,l}}}}{{{R_i}}} + {C_a}{m_{i,l}}({T_s} - {T_{i,l}})+q_{i,l},
\end{align}
where zone parameters $R_i$ and $C_i$ could be known by using model identification\cite{Goddard2014}\cite{Zhou2017}, and $q_{i,l}$ denotes external thermal disturbances of zone $i$ at time $l$ associated with zone occupancy states, lighting levels and so on, which can be measured by using IoT sensors or smart devices\cite{Minoli2017}.

By using finite-difference methods, \eqref{f_1} could be transformed into the discrete form over time intervals $\tau$ as follows,
\begin{align} \label{f_2}
T_{i,t+1}=d_iT_{i,t}+b_im_{i,t}(T_s-T_{i,t})+a_iT_{o,t}+\frac{\tau}{C_i}q_{i,t},
\end{align}
where $d_i=1-a_i$, $a_i=\frac{\tau}{R_iC_i}$, $b_i=\frac{\tau C_a}{C_i}$, $t\in [1,M]$ denotes the index of time intervals and $M$ is the total number of time intervals. Time interval $\tau$ is chosen so that outdoor temperature $T_{o,t}$, thermal disturbance $q_{i,t}$, electricity price $S_t$, and the most comfortable level in this interval could be regarded as the constants.

For indoor occupants, the thermal comfort range in zone $i$ could be described by
\begin{align} \label{f_3}
T_i^{\min}\leq T_{i,t}\leq T_i^{\max},~\forall~t,
\end{align}
where $T_i^{\min}$ and $T_i^{\max}$ denote the minimum and maximum indoor temperature, respectively.

The rate of the air supplied to each zone is controlled by a damper position in the VAV box, we have
\begin{align} \label{f_4}
m_i^{\min}\leq m_{i,t} \leq m_i^{\max},~\forall~i,t,
\end{align}
where $m_i^{\min}$ and $m_i^{\max}$ denote the minimum and maximum air supply rate of zone $i$.

Since the total air supply rate of the building is limited, we have the following constraint,
\begin{align} \label{f_5}
\sum\limits_i m_{i,t} \leq \overline{m},~\forall~t,
\end{align}
where $\overline{m}$ is assumed to be less than $\sum\nolimits_{i} m_i^{\max}$ so that \eqref{f_5} is nonredundant.

\subsection{Cost model}
The cost considered in this paper consists of three parts, namely thermal discomfort cost, energy cost associated with the supply fan, and energy cost associated with the cooling coil.

Similar to \cite{Constantopoulos1991}, we model the thermal discomfort cost of occupants at slot $t$ by
\begin{align} \label{f_6}
\Phi_{1,t}=\sum\limits_{i}\phi_i(T_{i,t+1}-T^{\text{ref}}_{i,t+1})^2,~\forall~t
\end{align}
where $\phi_i$ is the cost coefficient; $T^{\text{ref}}_{i,t+1}$ denotes the most comfortable temperature level of zone $i$ at slot $t+1$, and its value could be decided by the occupant at slot $t$.

Power consumption associated with the supply fan could be approximated by $\mu(\sum\nolimits_i m_{i,t})^3$\cite{Zhang2017}, where $\mu$ is a coefficient related to fan power consumption. Continually, the energy cost of the fan is given by
\begin{align} \label{f_7}
\Phi_{2,t}=\mu(\sum\nolimits_i m_{i,t})^3S_t\tau,~\forall~t.
\end{align}

The power consumption of the cooling coil could be represented by the following model\cite{Hao2017},
\begin{align} \label{f_8}
p_{t}=\frac{C_a\sum\nolimits_i m_{i,t}(T_m-T_s)}{\eta COP},
\end{align}
where $T_m=\gamma\frac{\sum\nolimits_i m_{i,t}T_{i,t}}{\sum\nolimits_i m_{i,t}}+(1-\gamma)T_{o,t}$ is the mixed air temperature, $\gamma\in[0,1]$ represents the damper position in the AHU. Substituting $T_m$ into \eqref{f_8}, $p_{t}$ could be rewritten as follows,
\begin{align} \label{f_9}
p_{t}=\sum\nolimits_i m_{i,t}\frac{C_a}{\eta COP}(\gamma T_{i,t}+(1-\gamma)T_{o,t}-T_s).
\end{align}
Continually, the energy cost associated with the cooling coil is given by
\begin{align} \label{f_10}
\Phi_{3,t}=p_{t}S_t\tau,~\forall~t.
\end{align}

\subsection{Problem formulation}
With the above-mentioned models, we formulate a problem to minimize the long-term total cost associated with the HAVC system as follows,
\begin{subequations}\label{f_11}
\begin{align}
(\textbf{P1})~&\min_{m_{i,t}}~\mathop {\lim\sup}\limits_{M \to \infty}\frac{1}{M-1}\sum\limits_{t=1}^{M-1} \mathbb{E}\{\sum\limits_{\ell=1}^{3}\Phi_{\ell,t}\}  \\
s.t.&~(2)-(5),
\end{align}
\end{subequations}
where $\mathbb{E}$ denotes the expectation operator, which acts on random electricity prices $S_t$, outdoor temperatures $T_{o,t}$, the most comfortable temperature level $T_{i,t+1}^{\text{ref}}$, and external thermal disturbances $q_{i,t}$; the decision variables of \textbf{P1} are $m_{i,t}$ ($\forall i,~t$).

\section{Algorithm Design}\label{s4}
\subsection{The proposed realtime algorithm}
There are several challenges involved in solving \textbf{P1}. Firstly, the future system parameters are unknown. Secondly, there are temporally coupled constraints \eqref{f_2} and spatially coupled constraints \eqref{f_5}. To deal with the above challenges, we intend to propose an algorithm based on the LOT framework, which has been widely used in energy management of data centers, microgrids, residential buildings, and smart homes. The key idea of a Lyapunov-based energy management algorithm is to buffer the power demand requests of flexible loads in energy queues when electricity prices are high and to serve such requests when electricity prices are low. Different from some loads (e.g., electric vehicles) with specific energy/power requirements, an HVAC has unknown power demand that is related to many factors, namely the most comfortable temperature level decided by occupants, the lower and upper bounds of indoor temperature, outdoor temperature, and external thermal disturbance. Thus, we need to redesign an algorithm to deal with the HVAC system in the commercial building. The key idea of the proposed algorithm CDRA is summarized as follows:
\begin{itemize}
  \item Constructing virtual queues associated with indoor temperatures of all zones.
  \item Obtaining the \emph{drift-plus-penalty} term according to the LOT framework.
  \item Minimizing the upper bound given in the right-hand-side of the \emph{drift-plus-penalty} term.
\end{itemize}

Based on the above idea, we can propose an online energy management algorithm without predicting any system parameters and knowing HVAC power demand in each time slot. Note that the purpose of constructing virtual queues is to guarantee the feasibility of constraints \eqref{f_3}. By stabilizing such queues, the proposed algorithm could operate without violating the constraints \eqref{f_3}. Specific proof can be found in Theorem 1.

To begin with, three assumptions are made about system parameters so that the system is controllable, i.e., (12)-(14). (12) implies that the temperature decrease of zone $i$ can be stopped by setting the minimum air rate $m_i^{\min}$ given minimum indoor temperature $T_i^{\min}$, minimum outdoor temperature $T_o^{\min}$, and minimum external disturbance $q_i^{\min}=\min_t q_{i,t}$. The intuition behind (13) is that the system control parameter $V$ defined in (23) should be positive. (14) is a sufficient but not necessary condition for the feasibility of the proposed algorithm, more details could be found in Appendix D.

\begin{figure*}[!t]
\normalsize
\setcounter{equation}{11}
\begin{align}\label{f_12}
&~~~~~~~~~~~~~~~~~~~~~~d_iT_i^{\min}+b_im_i^{\min}(T_s-T_i^{\min})+a_iT_o^{\min}+\frac{\tau}{C_i}q_i^{\min}\geq T_i^{\min},\\
&(T_i^{\max}-T_i^{\min})+a_i(T_o^{\min}-T_o^{\max})+\frac{\tau}{C_i}(q_i^{\min}-q_i^{\max})+b_i(m_i^{\max}(T_s-T_i^{\max})-m_i^{\min}(T_s-T_i^{\min}))>0,\\
&~~~~~~~~~~~~~~~~~~~~~~~~~~~~~~~\overline{m}\geq \sum\nolimits_{i}\frac{a_i(T_i^{\max}-T_o^{\max})-\frac{\tau}{C_i}q_i^{\max}}{b_i(T_s-T_i^{\min})}.
 \end{align}
\hrulefill
\vspace*{4pt}
\end{figure*}

\subsubsection{Constructing virtual queues}
To guarantee the feasibility of \eqref{f_3}, we define a virtual queue associated with indoor temperature $T_{i,t}$ as follows,
\begin{align} \label{f_13}
Q_{i,t}=T_{i,t}+\delta_i,
\end{align}
where $\delta_i$ ($\forall~i$) are constants, which are specified in Theorem 1. Then, $Q_{i,t+1}$ could be obtained as follows,
\begin{align} \label{f_14}
Q_{i,t+1}=T_{i,t+1}+\delta_i.
\end{align}

Substituting \eqref{f_2} and \eqref{f_13} into \eqref{f_14}, we have
\begin{align} \label{f_15}
&Q_{i,t+1}=(1-a_i)Q_{i,t}+b_im_{i,t}(T_s-T_{i,t}) \nonumber \\
&~~~~~~~~~~~+a_i(\delta_i+T_{o,t})+\frac{\tau}{C_i} q_{i,t}.
\end{align}

\subsubsection{Obtaining \emph{drift-plus-penalty} term}

To keep the virtual queues stable, we define a Lyapunov function below,
\begin{align} \label{f_16}
L_t=\frac{1}{2}\sum\limits_{i=1}^N Q_{i,t}^2.
\end{align}

Let $\boldsymbol{Q}_{t}$ be the vector $(Q_{1,t},Q_{2,t},\cdots,Q_{N,t})$. Then, we can compute the one-slot conditional Lyapunov drift as follows,
\begin{align} \label{f_17}
\Delta_t = \mathbb{E}\{ L_{t+1} - L_t|\boldsymbol{Q}_{t}\},
\end{align}
where the expectation is taken with respect to the randomness of electricity price, outdoor temperature, the most comfortable temperature level, and external thermal disturbance, as well as the chosen control decisions.

Taking \eqref{f_15} into consideration, we have
 \begin{align}\label{f_18}
 &L_{t+1}-L_t=\frac{1}{2}\sum\limits_{i=1}^N (Q_{i,t+1}^2-Q_{i,t}^2), \\
 &~~~~~~~~~~~\leq\frac{1}{2}\sum\limits_{i=1}^N (Q_{i,t+1}^2-(1-a_i)Q_{i,t}^2),\nonumber \\
 &~~~~~~~~~~~\leq \frac{1}{2}\sum\limits_{i=1}^N B_i+\sum\limits_{i=1}^N (1-a_i)Q_{i,t}b_i(T_s-T_{i,t})m_{i,t},\nonumber
 \end{align}
where $B_i=\Big(b_im_i^{\max}(T_s-T_i^{\max})^2+a_i(|\delta_i|+T_o^{\max})+\frac{\tau}{C_i}q_i^{\max}\Big)^2+2(1-a_i)(|\delta_i|+T_i^{\max})(a_i(|\delta_i|+T_o^{\max})+\frac{\tau}{C_i}q_i^{\max})$.

By adding a function of the expected total cost over one slot to \eqref{f_17}, we can obtain the \emph{drift-plus-penalty} term as follows,
\begin{align}\label{f_19}
&\Delta Y_t=\Delta_t + V\mathbb{E}\{\sum\limits_{\ell=1}^{3}\Phi_{\ell,t}|\boldsymbol{Q}_{t}\} \nonumber \\
&\leq \frac{1}{2}\sum\limits_{i=1}^N B_i+\mathbb{E}\{\sum\limits_{i=1}^N (1-a_i)Q_{i,t}b_i(T_s-T_{i,t})m_{i,t}|\boldsymbol{Q}_{t}\} \nonumber \\
&~~~+V\mathbb{E}\{\sum\limits_{\ell=1}^{3}\Phi_{\ell,t}|\boldsymbol{Q}_{t} \},
\end{align}
where $V$ is a positive control parameter to implement the tradeoff between queue stability and total cost minimization, and its value could be decided by (23).

\subsubsection{Minimizing the upper bound}
Since the key idea of the Lyapunov-based algorithm is to minimize the upper bound given in
the right-hand-side of the \emph{drift-plus-penalty} term. Then, we can propose a realtime HVAC control algorithm as in Algorithm 1.

\begin{algorithm}[h]
\caption{: Realtime HVAC Control Algorithm}
\begin{algorithmic}[1]
\STATE \textbf{For} each slot $t$ \textbf{do}\\
\STATE At the beginning of slot $t$, observe $\boldsymbol{Q}_t$,$T_{o,t}$, $S_t$, $T_{i,t+1}^{\text{ref}}$, and $q_{i,t}$; \\
\STATE Choose $m_{i,t}$ as the solution to \textbf{P2}:
\STATE (\textbf{P2})~$\min~\sum\limits_{i=1}^N (1-a_i)Q_{i,t}b_i(T_s-T_{i,t})m_{i,t}+V\sum\limits_{\ell=1}^{3}\Phi_{\ell,t}$  \\
\STATE ~~~~~~s.t.~(4),~(5),
\STATE Update $Q_{i,t}$ according to \eqref{f_15};
\STATE \textbf{End}
\end{algorithmic}
\end{algorithm}

\subsection{Solution to \textbf{P2}}
Since there are couplings among $m_{i,t}$ in the item $V\Phi_{2,t}$, \textbf{P2} cannot be solved easily. To make the problem tractable, $V\Phi_{2,t}$ is approximated by one of its upper bounds $V\mu S_t\tau N \overline{m}\sum\nolimits_i m_{i,t}^2$ using a Cauchy-Schwarz inequality\footnote{https://en.wikipedia.org/wiki/Cauchy-Schwarz\_inequality} since $(\sum\nolimits_i m_{i,t})^3\leq \overline{m}(\sum\nolimits_i m_{i,t})^2\leq N \overline{m}\sum\nolimits_i m_{i,t}^2$. Then, \textbf{P2} could be transformed into \textbf{P3} as follows,

\begin{subequations}\label{f_20}
\begin{align}
(\textbf{P3})~&\min_{m_{i,t}}~\sum\nolimits_i \Big((1-a_i)Q_{i,t}b_i(T_s-T_{i,t})m_{i,t} \nonumber\\
&~~~~~~~~+V\phi_i (T_{i,t+1}-T^{\text{ref}}_{i,t+1})^2 \nonumber \\
&~~~~~~~~+Vg_{i,t}m_{i,t}+V\mu S_t\tau N \overline{m} m_{i,t}^2\Big) \\
&s.t.~(4)-(5),
\end{align}
\end{subequations}
where $g_{i,t}=S_t\tau\frac{C_a}{\eta COP}(\gamma T_{i,t}+(1-\gamma)T_{o,t}-T_s)$.

\begin{figure}[!htb]
\centering
\includegraphics[scale=0.58]{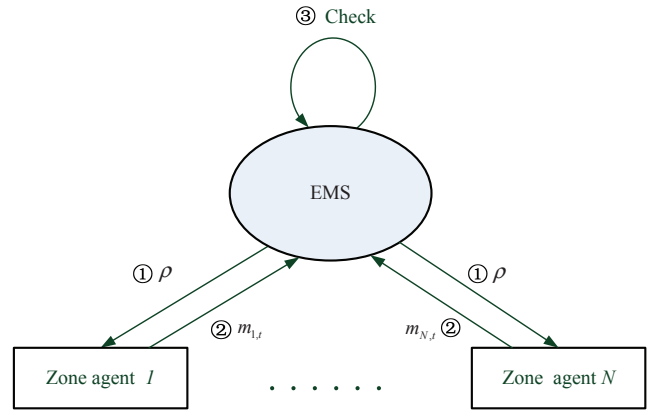}
\caption{Distributed implementation of the solution to \textbf{P3}.} \label{fig_2}
\end{figure}

In Appendix A, a solution based on binary search is adopted for \textbf{P3}. However, if the solution is implemented by a central Energy Management System (EMS) of the commercial building (e.g., a hotel), the concern associated with user privacy would be incurred. For example, if zone $i$ has unchanged $T_{i,t+1}^{\text{ref}}$ over several hours or very small $q_{i,t}$, it is probably that there is no occupant in zone $i$. Consequently, thieves may intrude into zone $i$ for stealing. To avoid the transmission of $T_{i,t+1}^{\text{ref}}$ and $q_{i,t}$, we provide a distributed implementation for the solution to \textbf{P3}. In the distributed implementation, $T_{i,t+1}^{\text{ref}}$ and $q_{i,t}$ are measured locally and are used to compute $m_{i,t}$. Then, $m_{i,t}$ is returned to the EMS for checking. The specific procedure could be found in Fig.~\ref{fig_2}, where three steps executed in one iteration of the solution are shown. Firstly, central EMS broadcasts $\rho_s$ (i.e., the value of $\rho$ at iteration $s$, and $\rho_1=0$) to all agents of zones. Then, each agent decides the value of $m_{i,t}$ according to \eqref{f_a2} and sends $m_{i,t}$ back to the central EMS. Finally, the central EMS checks the termination condition, i.e., $\sum\nolimits_i m_{i,t}<\overline{m}$ for $s=1$, while $\sum\nolimits_i m_{i,t}=\overline{m}$ for $s>1$. Compared with the centralized solution, the distributed solution has lower computation complexity (i.e., $\mathcal{O}(N_{\text{iter}})$, where $N_{\text{iter}}$ is the total iteration number) and offers high scalability with the increase of zone number.

\subsection{Algorithm feasibility}
Comparing the constraints of \textbf{P1} with those of \textbf{P2}, it can be observed that constraints \eqref{f_2} and \eqref{f_3} are neglected in \textbf{P2}. Due to the update of $Q_{i,t}$ ($\forall~i$) according to \eqref{f_15} in Algorithm 1, the constraint \eqref{f_2} could be satisfied by CDRA. To show the feasibility of CDRA to the original problem \textbf{P1}, we need to prove that \eqref{f_3} holds under CDRA.

As shown in \eqref{f_2}, the indoor temperature of zone $i$ is related to the air supply rate of each zone (i.e., $m_{i,t}$). To prove that \eqref{f_3} holds under CDRA, we first provide a Lemma about $m_{i,t}$ as follows.

\textbf{\emph{Lemma 1.}}
The optimal decision $m_{i,t}^\ddag$ of \textbf{P3} has the following properties ($m_{i,t}^*$ is defined as in \eqref{f_a1}),
\begin{enumerate}
  \item If $m_{i,t}^*<m_{i}^{\min}$, we have $m_{i,t}^\ddag=m_i^{\min}$.
  \item If $m_{i,t}^*>m_i^{\max}$, we have $m_{i,t}^\ddag\leq m_i^{\max}$.
\end{enumerate}
\begin{IEEEproof}
See Appendix B.
\end{IEEEproof}

Since $m_{i,t}^*$ is related to the value of $Q_{i,t}$, we can obtain the optimal decision information by checking the length of virtual queue $Q_{i,t}$ according to Lemma 1. Accordingly, we can obtain Lemma 2 as follows.

\textbf{\emph{Lemma 2.}} The optimal decision $m_{i,t}^\ddag$ of \textbf{P3} has the following properties (Definitions of $Q_{i}^a$ and $Q_{i}^b$ could be found in Appendix C),
\begin{enumerate}
  \item If $Q_{i,t}<Q_{i}^a$, $m_{i,t}^\ddag=m_i^{\min}$.
  \item If $Q_{i,t}>Q_{i}^b$, $m_{i,t}^\ddag\leq m_i^{\max}$.
\end{enumerate}
\begin{IEEEproof}
See Appendix C.
\end{IEEEproof}

Based on Lemma 2, we can prove the feasibility of CDRA as shown in Theorem 1 by considering three cases of $Q_{i,t}$, i.e., $[T_i^{\min}+\delta_i,~Q_{i}^a)$, $[Q_{i}^a,~Q_{i}^b]$, and $(Q_{i}^b,~T_i^{\max}+\delta_i]$.

\textbf{\emph{Theorem 1}}
Suppose the initial temperature level of zone $i$ $T_{i,0}\in [T_i^{\min},~T_i^{\max}]$, then, implementing CDRA with fixed parameters $V \in (0, V^{\max}]$ and $\delta_i \in [\delta_i^{\min},~\delta_i^{\max}]$, we have $T_{i,t} \in [T_{i}^{\min},~T_{i}^{\max}]$ for all slots (i.e., \eqref{f_3} could be satisfied under the proposed algorithm), where
\begin{align}
&V^{\max}=\min_i{\frac{\hbar_i}{\upsilon_i}},\\
&\delta_i^{\min}=\frac{\kappa_i^{\min}}{(1-a_i)}, \\
&\delta_i^{\max}=\frac{\kappa_i^{\max}}{(1-a_i)},
\end{align}
where $\hbar_i=(T_i^{\max}-T_i^{\min})+a_i(T_o^{\min}-T_o^{\max})+\frac{\tau}{C_i}(q_i^{\min}-q_i^{max})+b_i(m_i^{\max}(T_s-T_i^{\max})-m_i^{\min}(T_s -T_i^{\min}))$, $\upsilon_i=2\phi_i(T_i^{\text{refmax}}+d_iT_i^{\max}+a_iT_o^{\max}+\frac{\tau}{C_i}q_i^{\max}+b_i(T_i^{\max}-T_s)m_i^{\max})+\frac{g_i^{\max}+2m_i^{\max}\mu S^{\max}\tau N\overline{m}}{b_i(T_i^{\min}-T_s)}-\frac{g_i^{\min}+2m_i^{\min}\mu S^{\min}\tau N\overline{m}}{b_i(T_i^{\max}-T_s)}$, $\delta_i^{\min}$, $\kappa_i^{\min}=2V\phi_i(T_i^{\text{refmax}}+b_i(T_i^{\max}-T_s)m_i^{\max})+(Vg_{i}^{\max}+2V\mu \tau N\overline{m} m_i^{\max}S^{\max})/(b_i(T_i^{\min}-T_s))+b_im_i^{\min}(T_s-T_i^{\min})+a_iT_o^{\max}+\frac{\tau}{C_i}q_i^{\max}-T_i^{\max}$, $\kappa_i^{\max}=-2V\phi_i(d_iT_i^{\max}+a_iT_o^{\max}+\frac{\tau}{C_i}q_i^{\max})+(Vg_{i}^{\min}+2V\mu \tau N\overline{m} m_i^{\min}S^{\min})/(b_i(T_i^{\max}-T_s))+b_im_i^{\max}(T_s-T_i^{\max})+a_iT_o^{\min}+\frac{\tau}{C_i}q_i^{\min}-T_i^{\min}$.

\begin{IEEEproof}
See Appendix D.
\end{IEEEproof}

\subsection{Performance guarantee}
CDRA has the following performance guarantee as shown in Theorem 2, i.e., approaching to the optimal objective value of \textbf{P1} with an adjustable gap.

\textbf{\emph{Theorem 2}}
If electricity price $S_t$, outdoor temperature $T_{o,t}$, the most comfortable temperature level $T_{i,t+1}^{\text{ref}}$, and external thermal disturbance $q_{i,t}$ are i.i.d. over slots, CDRA has performance guarantee as follows, i.e., $\mathop {\lim\sup}\limits_{M \to \infty}\frac{1}{M-1}\sum\nolimits_{t=1}^{M-1} \mathbb{E}\{\sum\nolimits_{\ell=1}^{3}\Phi_{\ell,t}\}\leq y_1+\frac{\Theta}{V}$, where $y_1$ is the optimal objective value of \textbf{P1}, $\Theta=\frac{1}{2}\sum\nolimits_{i=1}^N B_i+\sum\nolimits_{i=1}^N (1-a_i)(T_i^{\max}+|\delta_i|)T_s(\frac{a_i(T_i^{\max}-T_o^{\min})-\frac{\tau}{C_i}q_i^{\min}}{(T_s+T_i^{\min})})$.

\begin{IEEEproof}
See Appendix E.
\end{IEEEproof}

\section{Performance Evaluation}\label{s5}

\begin{figure}
\centering
\subfigure[Retail electricity price]{
\begin{minipage}[b]{0.4\textwidth}
\includegraphics[width=1\textwidth]{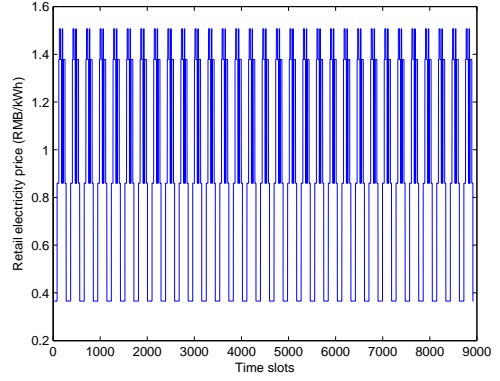}
\end{minipage}
}\\
\subfigure[Outdoor temperature]{
\begin{minipage}[b]{0.4\textwidth}
\includegraphics[width=1\textwidth]{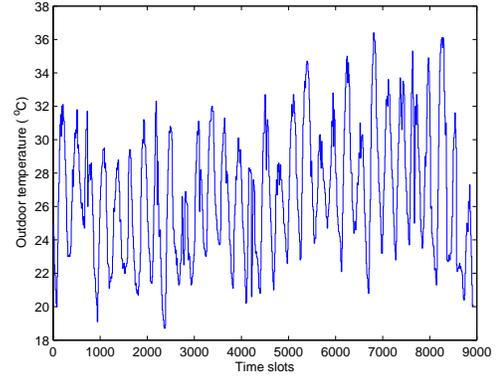}
\end{minipage}
}
\caption{Real-world traces used in simulations.} \label{fig_3}
\end{figure}

\subsection{Simulation setup}

We consider a time horizon with $M=8928$ intervals and the length of a time interval $\tau$ is 5 minutes (note that outdoor temperature and thermal disturbance vary at a time-scale of minutes in practice\cite{Zhang2017}, while electricity prices and the most comfortable temperature levels usually vary at a time-scale of hours. In other words, the length of the time horizon is one month with 31 days. $N=4$, $\gamma=0.95$, $T_i^{\min}=18^oC$, $V=V^{\max}$, $\delta_i=\delta_i^{\min}$. Main parameters associated with each zone and the HVAC system are configured as follows: $T_s=12.8^oC$\cite{Zhang2017}, $C_a=1.012J/g/^oC$\cite{Zhang2017}, $R_i=[0.0053, 0.0060, 0.0063, 0.0067]^oC/W$\cite{Hao2012}, $C_i=[550000, 570000, 590000, 620000]J/^oC$\cite{Hao2012}, $\mu=2\times 10^{-6}W/(g/s)^3$\cite{Zhang2017}, $m_i^{\min}=0g/s$, $m_i^{\max}=450g/s$\cite{Zhang2017}, $\overline{m}=1400g/s$, $\eta=0.8879$\cite{Hao2017}, $COP=5.9153$\cite{Hao2017}. For electricity price information, we adopt the hourly retail commercial electricity price associated with Beijing city of China in July of 2017\footnote{http://www.95598.cn/static/html//person/sas/es//PM06003001\_786.shtml}. Due to the lack of hourly outdoor temperature traces in Beijing city, we adopt the hourly outdoor temperature trace related to Edmonton city of Canada in July of 2017\footnote{http://www.theweathernetwork.com} since retail electricity prices are independent of outdoor temperatures in Beijing. Moreover, considering temperature differences of two cities, we raise outdoor temperatures in Edmonton city by $8^oC$ so that the obtained temperature range (i.e., [$18.7^oC\backsim 36.4^oC$]) is close to that in Beijing (i.e., [$19.0^oC\backsim 36.0^oC$]). In addition, we assume that the most comfortable temperature levels in an hour at all zones follow discrete uniform distributions with parameters 21 and 23 ($^oC$)\cite{Zhang2017}. Moreover, external thermal disturbances in a time interval follows a uniform distribution with parameters 0.1 and 0.2 ($W$)\cite{Zhang2017}.

The simulations are conducted using MATLAB 2013a on a computer with 4 GB memory and a Core i7 CPU of frequency 2.4 GHz. For performance comparisons, three baselines are adopted as follows.
\begin{itemize}
  \item B1: this baseline intends to maintain the most comfortable temperature level $T_{i,t+1}^{\text{ref}}$ for all zones as adopted in \cite{Hao2017}. Moreover, when the total required air supply rate $\sum\nolimits_i m_{i,t}$ is greater than $\overline{m}$, we set the actual air supply rate of each zone $m_{i,t}^*$ as $\frac{m_{i,t}}{\sum\nolimits_i m_{i,t}}\overline{m}$ for the proportional fairness.
  \item B2: similar to the greedy algorithm in \cite{SunTSG2016}, this baseline intends to greedily minimize the current total cost $\sum\nolimits_{\ell=1}^{3}\Phi_{\ell,t}$ in each time slot $t$ without violating the temperature limits as described in \eqref{f_3}.
  \item Modified CDRA (MCDRA): this baseline is the same as CDRA except that $\delta_i=[-(\frac{\phi_i}{\phi_i^{\max}}T_{i,t+1}^{\text{ref}}+(1-\frac{\phi_i}{\phi_i^{\max}})(-\delta_i^{\min}))]^\mathcal{P}$, where $\phi_i^{\max}$ is the maximum cost coefficient, $[\dagger]^\mathcal{P}=\max(\delta_i^{\min},\min(\delta_i^{\max},\dagger))$. The intuition behind the above setting is that $-\delta_i$ represents our expected indoor temperature and it should be close to $T_{i,t+1}^{\text{ref}}$ if $\phi_i$ equals to $\phi_i^{\max}$.
\end{itemize}

\begin{figure}
\centering
\subfigure[$Q_{i,t}$]{
\begin{minipage}[b]{0.4\textwidth}
\includegraphics[width=1\textwidth]{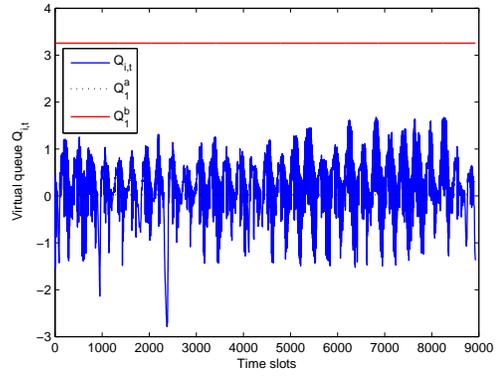}
\end{minipage}
}\\
\subfigure[Air supply rate]{
\begin{minipage}[b]{0.4\textwidth}
\includegraphics[width=1\textwidth]{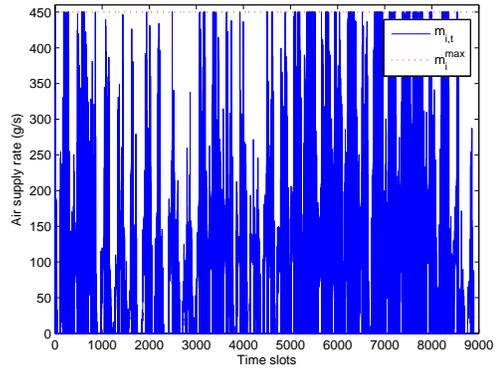}
\end{minipage}
}\\
\subfigure[Total air supply rate]{
\begin{minipage}[b]{0.4\textwidth}
\includegraphics[width=1\textwidth]{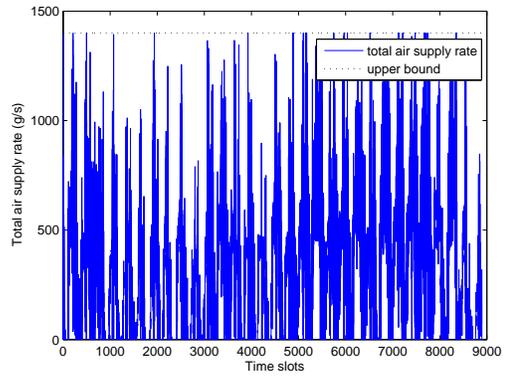}
\end{minipage}
}\\
\subfigure[Indoor temperature ($T_i^{\max}=26^oC$)]{
\begin{minipage}[b]{0.4\textwidth}
\includegraphics[width=1\textwidth]{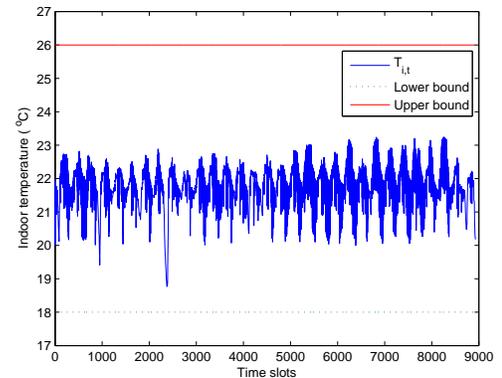}
\end{minipage}
}
\caption{The feasibility of the proposed algorithm ($i=1$).} \label{fig_4}
\end{figure}

\begin{figure}
\centering
\subfigure[Energy cost]{
\begin{minipage}[b]{0.4\textwidth}
\includegraphics[width=1\textwidth]{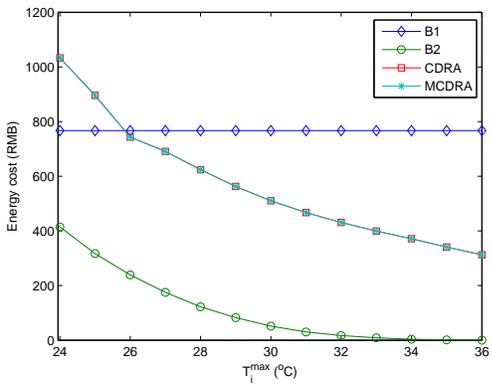}
\end{minipage}
}\\
\subfigure[ATD]{
\begin{minipage}[b]{0.4\textwidth}
\includegraphics[width=1\textwidth]{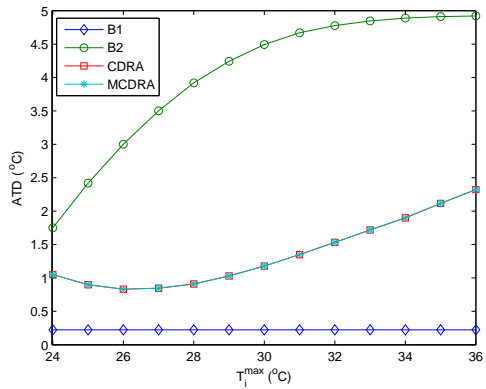}
\end{minipage}
}\\
\subfigure[Indoor temperature ($T_i^{\max}=30^oC$)]{
\begin{minipage}[b]{0.4\textwidth}
\includegraphics[width=1\textwidth]{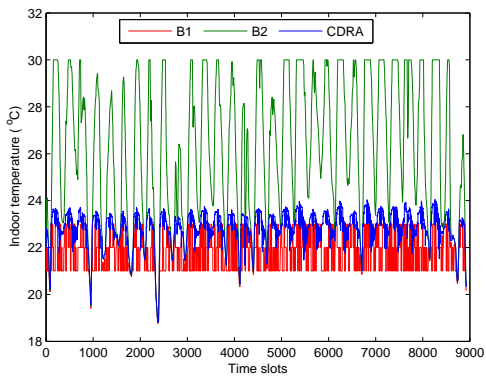}
\end{minipage}
}\\
\subfigure[Average indoor temperature]{
\begin{minipage}[b]{0.4\textwidth}
\includegraphics[width=1\textwidth]{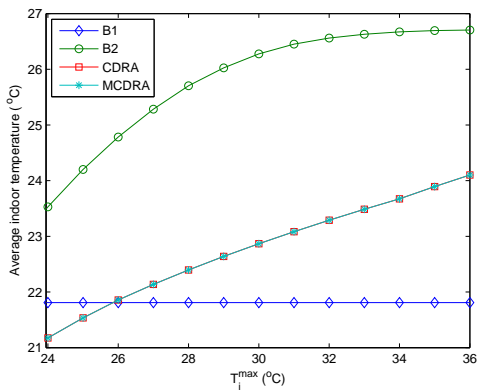}
\end{minipage}
}
\caption{The impact of $T_i^{\max}$ ($\phi_i=0$)}. \label{fig_5}
\end{figure}

\begin{figure}
\centering
\subfigure[Total cost]{
\begin{minipage}[b]{0.4\textwidth}
\includegraphics[width=1\textwidth]{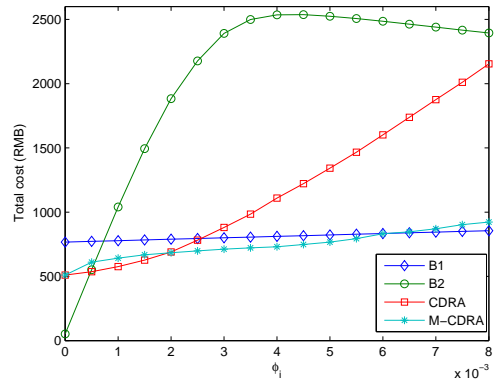}
\end{minipage}
}
\subfigure[Energy cost]{
\begin{minipage}[b]{0.4\textwidth}
\includegraphics[width=1\textwidth]{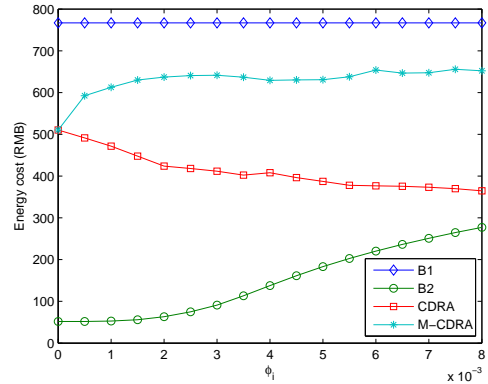}
\end{minipage}
}
\subfigure[Thermal discomfort cost]{
\begin{minipage}[b]{0.4\textwidth}
\includegraphics[width=1\textwidth]{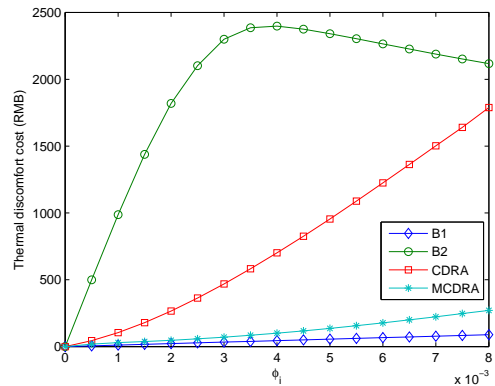}
\end{minipage}
}
\subfigure[ATD]{
\begin{minipage}[b]{0.4\textwidth}
\includegraphics[width=1\textwidth]{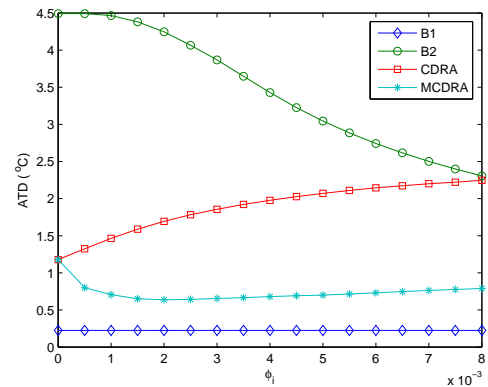}
\end{minipage}
}
\caption{The impact of $\phi_i$ ($T_i^{\max}=30^oC$)}. \label{fig_6}
\end{figure}

\subsection{Simulation results}
\subsubsection{Algorithm feasibility}
The algorithm feasibility of CDRA could be illustrated by Fig.~\ref{fig_4}, where indoor temperature $T_{i,t}$ and air supply rate $m_{i,t}$ always fluctuate within their respective normal ranges\footnote{Just the results associated with zone 1 are shown for brevity.}. By taking Figs.~\ref{fig_4}(a) and (b) into consideration, it can be found that $m_{i,t}=m_i^{\min}$ if $Q_{i,t}<Q_{i}^a$, which verifies the Lemma 2. Moreover, it can be observed that the total air supply rate is always less than $\overline{m}$. Thus, the constraints (3)-(5) could be guaranteed by CDRA. In addition, updating $Q_{i,t}$ in the Algorithm 1 according to \eqref{f_15} means that \eqref{f_2} could also be guaranteed. In summary, all constraints in the original problem \textbf{P1} could be satisfied by CDRA.

\subsubsection{The impact of $T_i^{\max}$}
In this subsection, we evaluate the impact of $T_i^{\max}$ on the total cost under different schemes by fixing the value of $\phi_i$. To be specific, $\phi_i$ is fixed to be zero and $T_i^{\max}$ varies from $24^oC$ to $36^oC$. At this time, MCDRA is equivalent to CDRA and total cost is equal to energy cost. As shown in Figs.~\ref{fig_5}(a)-(b), larger $T_i^{\max}$ contributes to reducing energy cost with the sacrifice of ATD, which is defined as the Average Temperature Deviation from the most comfortable temperature level $T^{\text{ref}}_{t+1}$, i.e., $\text{ATD}=\frac{1}{N(T-1)}\sum\nolimits_{i=1}^{N}\sum\nolimits_{t=0}^{T-2}|T_{i,t+1}-T^{\text{ref}}_{i,t+1}|$. The reason is obvious since larger temperature range means more opportunities\footnote{Due to the building system inertia, the HVAC system would consume less energy in later time slots with high electricity prices if it consumes more energy in the current time slot with low electricity price. Consequently, lower energy cost and larger volatility of indoor temperatures are incurred.} for the cost reduction. When the allowed ATD is $1^oC$, CDRA (or MCDRA) could reduce energy cost by 26.6\% when compared with B1. Though B2 achieves the lowest energy cost, the corresponding ATD is also the largest. The reason is that B2 intends to maintain indoor temperature around $T_i^{\max}$ so that the current total cost could be minimized, which is illustrated by Fig.~\ref{fig_5}(c). In Fig.~\ref{fig_5}(b), the \text{ATD} curve under CDRA does not monotonically increase with the increase of $T_i^{\max}$. The reason for such tendency is that the average indoor temperature under CDRA (MCDRA) first approaches the most comfortable temperature range and then departs from that range as shown in Fig.~\ref{fig_5}(d). In next subsection, we would evaluate the impact of $\phi_i$ on the total cost since occupants in the zones usually care about thermal comforts in practice (i.e., $\phi_i>0$).

\subsubsection{The impact of $\phi_i$}
$\phi_i$ reflects the relative importance of thermal discomfort with respect to energy cost. In extreme cases, $\phi_i=\infty$ means that the temperature deviation from the most comfortable temperature level decided by the occupant of zone $i$ is not permitted, while $\phi_i=0$ means that the occupant in zone $i$ does not care about the change of indoor temperature as long as the indoor temperature limits \eqref{f_3} could be satisfied. The impact of $\phi_i$ on the performances of all schemes could be found in Fig.~\ref{fig_6}, where CDRA and MCDRA achieve the lowest total cost when $\phi_i\in [0.0005,~0.002]$ and $\phi_i\in (0.002,~0.006]$, respectively. When $\phi_i> $0.006, the best way of controlling HVAC is to maintain the most comfortable temperature level. In addition, there is a flexible tradeoff between energy cost and ATD under CDRA (or MCDRA) by observing Figs.~\ref{fig_6}(b) and (c), e.g., when the allowed $\text{ATD} \in(0.63,~2.25)^oC$ and $T_i^{\max}=30^oC$, the relative energy cost reduction is varying from 16.92\% to 52.43\%. In contrast, the flexible tradeoff can not be supported by B1. Though there is a flexible tradeoff under B2, its overall performance (i.e., the total cost) is worse than CDRA and MCDRA when $\phi_i\geq 0.001$ as shown in Fig.~\ref{fig_6}(a). Thus, the proposed algorithm offers an effective way of controlling HVAC systems for commercial buildings when energy cost and thermal discomfort are considered jointly.

\section{Conclusions and Future Work}\label{s6}
In this paper, we investigated the problem of minimizing the time average expected total cost related to an HVAC system in a commercial building with the consideration of uncertainties in electricity price, outdoor temperature, the most comfortable temperature level, and external thermal disturbance. Then, we proposed a cost-ware distributed realtime algorithm (CDRA) to control the HVAC system without requiring the prediction of future system parameters and knowing their stochastic information. Moreover, CDRA protects user privacy from being exploited by attackers and offers high algorithmic scalability. Extensive simulation results based on real-world traces showed the effectiveness of the proposed algorithm. In the future, we plan to design a cost-efficient HVAC control algorithm for a commercial building by jointly adjusting the air supply rate of each zone and the damper position in the AHU, since different zones usually have varied ventilation requirements due to different floor areas and occupancy states. In addition, a natural extension of this work is to design a distributed realtime HVAC control algorithm for a commercial building with several shared spaces (e.g., large meeting rooms), where different occupants have different temperature preferences.

\appendices

\section{The solution to \textbf{P3}}
\begin{IEEEproof}
Let the first derivative of $\Upsilon$ with respect to $m_{i,t}$ be zero, we have
 \begin{align} \label{f_a1}
m_{i,t}^*=\frac{b_i(T_s-T_{i,t})h_{i,t}-Vg_{i,t}}{2V\mu S_t\tau N\overline{m}+2V\phi_i b_i^2(T_s-T_{i,t})^2},
\end{align}
where $\Upsilon$ is the objective function of \textbf{P3}, $h_{i,t}=2V\phi_i (T_{i,t+1}^\text{ref}-d_iT_{i,t}-a_iT_{o,t}-\frac{\tau}{C_i}q_{i,t})-(1-a_i)Q_{i,t}$. Since $m_i^{\min}\leq m_{i,t}\leq m_i^{\max}$, the optimal value of $m_{i,t}$ (i.e., $m_{i,t}^\diamond$) would be $\max(m_i^{\min},\min(m_i^{\max},m_{i,t}^*))$ if \eqref{f_5} is neglected. If $\sum\nolimits_{i=1}^N m_{i,t}^\diamond\leq \overline{m}$, the optimal solution to \textbf{P3} is found. Otherwise, using KKT optimality conditions, the optimal solution of \textbf{P3} is given as follows,
\begin{align} \label{f_a2}
m_{i,t}^\ddag=\max(m_i^{\min},\min(m_i^{\max},m_{i,t}^\dag));
\end{align}
where $m_{i,t}^\dag=\frac{b_i(T_s-T_{i,t})h_{i,t}-Vg_{i,t}-\rho}{2V\mu S_t\tau N\overline{m}+2V\phi_i b_i^2(T_s-T_{i,t})^2}$, $\rho$ is a non-negative dual variable related to \eqref{f_5}. With the increase of $\rho$, $m_{i,t}^\dag$ is gradually reduced. According to the complementary slackness condition in KKT conditions, the optimal $\rho$ is a value that leads to $\sum\nolimits_{i}m_{i,t}^\ddag=\overline{m}$. To find the optimal $\rho$, binary search could be used\cite{Liang2016TSG,LiangIoT2016} and corresponding algorithmic complexity is $\mathcal{O}(N_{\text{iter}}N)$, where $N_{\text{iter}}$ is the total iteration number.
\end{IEEEproof}

\section{Proof of Lemma 1}
\begin{IEEEproof}
Note that the objective function of \textbf{P3} $\Upsilon$ is a decomposable quadratic function over variable $m_{i,t}$ and the corresponding parabolas open upwards, we have $m_{i,t}=m_i^{\min}$ if $m_{i,t}^*<m_i^{\min}$, and $m_{i,t}=m_i^{\max}$ if $m_{i,t}^*>m_i^{\max}$. When considering constraints \eqref{f_4} and \eqref{f_5} simultaneously, the optimal $m_{i,t}$ is described by \eqref{f_a2}. Since $\rho\geq 0$, we have $m_{i,t}^\ddag=m_i^{\min}$ if $m_{i,t}^*<m_i^{\min}$, and $m_{i,t}^\ddag\leq m_i^{\max}$ if $m_{i,t}^*>m_i^{\max}$.
\end{IEEEproof}

\section{Proof of Lemma 2}
\begin{IEEEproof}
According to \eqref{f_a1}, $m_{i,t}^*$ would be smaller than $m_i^{\min}$ when $Q_{i,t}<\frac{\chi_{i,t}}{b_i(1-a_i)(T_s-T_{i,t})}$, where $\chi_{i,t}=b_i(T_s-T_{i,t})(2V\phi_i (T_{i,t+1}^\text{ref}-d_iT_{i,t}-a_iT_{o,t}-\frac{\tau}{C_i}q_{i,t}))-vg_{i,t}-m_i^{\min}(2V\mu S_t\tau N \overline{m}+2V\phi_i b_i^2(T_s-T_{i,t})^2)$. Since the minimum value of $\frac{\chi_{i,t}}{b_i(1-a_i)(T_s-T_{i,t})}$ (i.e., $Q_{i}^a$) is $\frac{Vg_i^{\min}+2m_i^{\min}V\mu S^{\min}\tau N \overline{m}}{b_i(1-a_i)(T_i^{\max}-T_s)}-\frac{2V\phi_i(d_iT_i^{\max}+a_iT_o^{\max}+\frac{\tau}{C_i}q_i^{\max})}{(1-a_i)}$, we have $m_{i,t}^*<m_i^{\min}$ when $Q_{i,t}<Q_{i}^a$. According to Lemma 1, we have $m_{i,t}^\ddag=m_i^{\min}$. Similarly, when $Q_{i,t}>Q_{i}^b$ ($Q_{i}^b=\frac{Vg_i^{\max}+2m_i^{\max}V\mu S^{\max}\tau N\overline{m}}{b_i(1-a_i)(T_i^{\min}-T_s)}+\frac{2(T_i^{\max}-T_s)m_i^{\max}V\phi_i b_i}{(1-a_i)}+\frac{2T_i^{\text{refmax}}V\phi_i}{(1-a_i)}$), we have $m_{i,t}^*>m_i^{\max}$. According to Lemma 1, we have $m_{i,t}^\ddag\leq m_i^{\max}$.
\end{IEEEproof}

\section{Proof of Theorem 1}
We will prove that the above inequalities are satisfied for all time slots by using mathematical induction method.
Since $T_i^{\min}\leq T_{i,0}\leq T_i^{\max}$, the above inequalities hold for $t$=0. Suppose the above-mentioned inequalities hold for the time slot $t$, we should verify that they still hold for the time slot $t$+1. Specifically, we consider three cases as follows.
\begin{itemize}
  \item If $Q_{i}^b <Q_{i,t}\leq T_i^{\max}+\delta_i$, then, $m_{i,t}^\ddag\leq m_i^{\max}$ according to Lemma 2. As a result, $T_{i,t+1}=d_iT_{i,t}+b_im_{i,t}^\ddag(T_s-T_{i,t})+a_iT_{o,t}+\frac{\tau}{C_i}q_{i,t}\geq d_i(Q_{i}^b-\delta_i)+b_im_i^{\max}(T_s-T_i^{\max})+a_iT_o^{\min}+\frac{\tau}{C_i}q_i^{\min}\geq T_i^{\min}$, where $\delta_i=\delta_i^{\max}$ is adopted. Moreover, $T_{i,t+1}\leq d_iT_i^{\max}+b_i m_{i,t}^\ddag(T_s-T_i^{\min})+a_iT_o^{\max}+\frac{\tau}{C_i}q_i^{\max}$. In order to ensure that $T_{i,t+1}\leq T_i^{\max}$ for all $i$, we have $m_{i,t}^\ddag\geq \frac{a_i(T_i^{\max}-T_o^{\max})-\frac{\tau}{C_i}q_i^{\max}}{b_i(T_s-T_i^{\min})}$. According to \eqref{f_5}, we have $\sum\nolimits_{i}m_{i,t}^\ddag\leq \overline{m}$. Consequently, $\overline{m}\geq \sum\nolimits_{i}\frac{a_i(T_i^{\max}-T_o^{\max})-\frac{\tau}{C_i}q_i^{\max}}{b_i(T_s-T_i^{\min})}$. Taking into account the truth that all zones have different physical properties, the above case would not happen simultaneously for all zones. Therefore, the above condition is sufficient but not necessary for the feasibility of the proposed algorithm.
  \item If $T_i^{\min}+\delta_i \leq Q_{i,t}<Q_{i}^a$, then, we have $m_{i,t}^\ddag=m_i^{\min}$ according to Lemma 2. As a result, $T_{i,t+1}=d_iT_{i,t}+b_im_{i,t}(T_s-T_{i,t})+a_iT_{o,t}+\frac{\tau}{C_i}q_{i,t}\leq d_i(Q_{i}^a-\delta_i)+b_im_i^{\min}(T_s-T_i^{\min})+a_iT_o^{\max}+\frac{\tau}{C_i}q_i^{\max}\leq T_i^{\max}$, where $\delta_i=\delta_i^{\min}$ is adopted. Moreover, $T_{i,t+1}\geq (d_i-b_im_i^{\min})T_i^{\min}+b_im_i^{\min}T_s+a_iT_o^{\min}+\frac{\tau}{C_i}q_i^{\min}\geq T_i^{\min}$, where (12) and $d_i\geq b_im_i^{\min}$ are adopted. Since $m_i^{\min}$ is usually close to zero due to the minimal ventilation requirement, the inequality $d_i\geq b_im_i^{\min}$ holds in practice.
  \item If $Q_{i}^a \leq Q_{i,t} \leq Q_{i}^b$, we have $T_{i,t+1}\leq d_i(Q_{i}^b-\delta_i)+b_im_i^{\min}(T_s-T_i^{\min})+a_iT_{o}^{\max}+\frac{\tau}{C_i}q_{i}^{\max}\leq T_i^{\max}$, where $\delta_i=\delta_i^{\min}$ is adopted. Similarly, we have $T_{i,t+1}\geq d_i(Q_{i}^a-\delta_i)+b_im_i^{\max}(T_s-T_i^{\max})+a_iT_{o}^{\min}+\frac{\tau}{C_i}q_{i}^{\min}\geq T_i^{\min}$, where $\delta_i=\delta_i^{\max}$ is adopted.
\end{itemize}
In summary, $T_i^{\min}\leq T_{i,t}\leq T_i^{\max}$ for any zone $i$ and any time slot $t$, which completes the proof.

\section{Proof of Theorem 2}
We first define some equations as follows for the convenience of analysis, i.e.,
$\overline{T_i}=\mathop {\lim \sup }\limits_{M \to \infty } \frac{1}{M-1}\sum\limits_{t=1}^{M-1} \mathbb{E}\{T_{i,t}\},$
$\overline{T_o}=\mathop {\lim \sup }\limits_{M \to \infty } \frac{1}{M-1}\sum\limits_{t = 1}^{M-1} \mathbb{E}\{T_{o,t}\},$
$\overline{q_i}=\mathop {\lim \sup }\limits_{M \to \infty } \frac{1}{M-1}\sum\limits_{t=1}^{M-1} \mathbb{E}\{q_{i,t}\}, $
$\overline{m_i}=\mathop {\lim \sup }\limits_{M \to \infty } \frac{1}{M-1}\sum\limits_{t = 1}^{M - 1} \mathbb{E}\{m_{i,t}\}$.

Then, based on the constraint \eqref{f_2}, we can obtain the following inequalities, i.e., $\frac{a_i(T_i^{\min}-T_o^{\max})-\frac{\tau}{C_i}q_i^{\max}}{b_i(T_s+T_i^{\max})}\leq \overline{m_i}\leq \frac{a_i(T_i^{\max}-T_o^{\min})-\frac{\tau}{C_i}q_i^{\min}}{b_i(T_s+T_i^{\min})}$.

Next, we consider the following optimization problem,
\begin{subequations}\label{f_a4}
\begin{align}
(\textbf{P4})~&\min~\mathop {\lim\sup}\limits_{M \to \infty}\frac{1}{M-1}\sum\limits_{t=1}^{M-1} \mathbb{E}\{\sum\limits_{\ell=1}^{3}\Phi_{\ell,t}\}  \\
s.t.&~(4),(5),\\
&\overline{m_i}\geq \frac{a_i(T_i^{\min}-T_o^{\max})-\frac{\tau}{C_i}q_i^{\max}}{b_i(T_s+T_i^{\max})},\\
&\overline{m_i}\leq \frac{a_i(T_i^{\max}-T_o^{\min})-\frac{\tau}{C_i}q_i^{\min}}{b_i(T_s+T_i^{\min})}.
\end{align}
\end{subequations}
Note that (2),(3) in \textbf{P1} are replaced by (28c),(28d) in \textbf{P4}. Since any feasible solution of \textbf{P1} is also feasible to \textbf{P4}, we have $y_2\leq y_1$, where $y_2$ and $y_1$ are the optimal objective values of \textbf{P4} and \textbf{P1}, respectively. Using the Theorem 4.5 in \cite{Neely2010}, a conclusion could be obtained similarly, i.e., if electricity price $S_t$, outdoor temperature $T_{o,t}$, the most comfortable temperature level $T_{i,t+1}^{\text{ref}}$, external thermal disturbance $q_{i,t}$ are i.i.d. over slots and \textbf{P4} is feasible, there exists a stationary, randomized policy that takes control decision $m_{i,t}^*$ purely as a function of current system observation parameters and provides the following performance guarantee, i.e., $\mathbb{E}\{\sum\nolimits_{\ell=1}^{3}\Phi_{\ell,t}^*\}\leq y_2$, $\mathbb{E}\{m_{i,t}^*\}\leq \frac{a_i(T_i^{\max}-T_o^{\min})-\frac{\tau}{C_i}q_i^{\min}}{b_i(T_s+T_i^{\min})}$. Continually, when using the proposed algorithm, we have
\begin{align} \label{f_th41}
&\Delta Y_t=\Delta_t + V\mathbb{E}\{\sum\limits_{\ell=1}^{3}\Phi_{\ell,t}|\boldsymbol{Q}_{t}\} \nonumber \\
&\leq \frac{1}{2}\sum\limits_{i=1}^N B_i+\mathbb{E}\{\sum\limits_{i=1}^N (1-a_i)Q_{i,t}b_i(T_s-T_{i,t})m_{i,t}^*|\boldsymbol{Q}_{t}\} \nonumber \\
&~~~+V\mathbb{E}\{\sum\limits_{\ell=1}^{3}\Phi_{\ell,t}^*|\boldsymbol{Q}_{t} \},\\
&\leq \frac{1}{2}\sum\limits_{i=1}^N B_i+Vy_2+\xi,\\
&\leq \Theta+Vy_1,
\end{align}
where $\xi$=$\sum\limits_{i=1}^N (1-a_i)(T_i^{\max}+|\delta_i|)T_s(\frac{a_i(T_i^{\max}-T_o^{\min})-\frac{\tau}{C_i}q_i^{\min}}{(T_s+T_i^{\min})})$, $\Theta=\frac{1}{2}\sum\nolimits_{i=1}^N B_i+\xi$, (29) holds due to that the proposed algorithm minimizes the upper bound given in the right-hand-side of the \emph{drift-plus-penalty} term over all other control strategies, including the optimal stationary and randomized control strategy; (30) is obtained by incorporating the results of a stationary, randomized control strategy associated with \textbf{P4}. By arranging the both sides of the above equations, we have $\mathbb{E}\{\Delta_t\} + V\mathbb{E}\{\sum\nolimits_{\ell=1}^{3}\Phi_{\ell,t}\} \leq \Theta+Vy_1$. Continually, we have $V {\sum\nolimits_{t=1}^{M-1} {\mathbb{E}\{\sum\nolimits_{\ell=1}^{3}\Phi_{\ell,t}\}}} \leq \Theta (M-1)+V(M-1)y_1 - \mathbb{E}\{L_{M-1}\}+\mathbb{E}\{L_1\}$. Dividing both side by $V(M-1)$, and taking a \text{lim sup} of both sides. Then, let $M \to \infty$, we have $\mathop {\lim \sup }\limits_{M \to \infty } \frac{1}{M-1}{\sum\nolimits_{t = 1}^{M-1} {\mathbb{E}\{\sum\nolimits_{\ell=1}^{3}\Phi_{\ell,t}\}}} \le y_1+\frac{\Theta}{V}$, which completes the proof.

\end{document}